\documentclass{sig-alternate-05-2015}
\usepackage{url}
\usepackage{latexsym}
\usepackage{color}
\usepackage{graphicx}
\usepackage{multirow}
\usepackage{enumitem}
\usepackage{changes}
\usepackage{float}
\usepackage{etoolbox}
\patchcmd{\maketitle}{\@copyrightspace}{}{}{}

\begin{document}






%
\conferenceinfo{WSDM Workshop on Mining Online Health Reports}{'17 Cambridge, UK}

\title{Feature Studies to Inform the Classification of Depressive Symptoms from Twitter Data for Population Health} 

%
%
%
%
%

\numberofauthors{3} 
%

%
%
\author{
\alignauthor
Danielle Mowery\\
       \affaddr{University of Utah}\\
       \affaddr{Biomedical Informatics}\\
       \affaddr{421 Wakara Way}\\
       \affaddr{Salt Lake City, UT, USA}\\
       \email{danielle.mowery@utah.edu}
\alignauthor
Craig Bryan\\
       \affaddr{University of Utah}\\
       \affaddr{Psychology}\\
       \affaddr{380 S 1530 E BEH S 502}\\
       \affaddr{Salt Lake City, UT, USA}\\
       \email{craig.bryan@utah.edu}
\alignauthor Mike Conway\\
       \affaddr{University of Utah}\\
       \affaddr{Biomedical Informatics}\\
       \affaddr{421 Wakara Way}\\
       \affaddr{Salt Lake City, UT, USA}\\
       \email{mike.conway@utah.edu}
}

\maketitle
\begin{abstract}
The utility of Twitter data as a medium to support population-level mental health monitoring is not well understood. In an effort to better understand the predictive power of supervised machine learning classifiers and the influence of feature sets for efficiently classifying depression-related tweets on a large-scale, we conducted two feature study experiments. In the first experiment, we assessed the contribution of feature groups such as lexical information (e.g., unigrams) and emotions (e.g., strongly negative) using a feature ablation study. In the second experiment, we determined the percentile of top ranked features that produced the optimal classification performance by applying a three-step feature elimination approach. In the first experiment, we observed that lexical features are critical for identifying \emph{depressive symptoms}, specifically for \emph{depressed mood} (-35 points) and for \emph{disturbed sleep} (-43 points). In the second experiment, we observed that the optimal F1-score performance of top ranked features in percentiles variably ranged across classes e.g., \emph{fatigue or loss of energy} (5th percentile, 288 features) to \emph{depressed mood} (55th percentile, 3,168 features) suggesting there is no consistent count of features for predicting depressive-related tweets. We conclude that simple lexical features and reduced feature sets can produce comparable results to larger feature sets. 

\end{abstract}


%
%



\section{Introduction}

In recent years, there has been a movement to leverage social medial
data to detect, estimate, and track the change in prevalence of
disease.  For example,  eating
disorders in Spanish language Twitter tweets \cite{Prieto:2014aa} and
influenza surveillance \cite{Collier:2011aa}. More recently, social
media has been leveraged to monitor social risks such as prescription
drug and smoking behaviors \cite{Myslin,Hanson,Chen} as well as a
variety of mental health disorders including suicidal ideation
\cite{DeChoudhury2016}, attention deficient hyperactivity disorder
\cite{coppersmith2015adhd} and major depressive disorder
\cite{DeChoudhury2014}. In the case of
major depressive disorder, recent efforts range from characterizing
linguistic phenomena associated with depression
\cite{coppersmith-dredze-harman:2014:W14-32} and its subtypes e.g.,
postpartum depression  \cite{DeChoudhury2016}, to identifying specific
depressive symptoms \cite{Cavazos-Rehg,Moweryc} e.g., depressed mood. However, more
research is needed to better understand the predictive power of
supervised machine learning classifiers and the influence of feature groups and feature sets for efficiently classifying depression-related
tweets to support mental health monitoring at the population-level \cite{Conway}.

This paper builds upon related works toward classifying Twitter tweets
representing symptoms of major depressive disorder by assessing
the contribution of lexical features (e.g.,  unigrams) and emotion
(e.g., strongly negative) to classification performance, and by applying methods to eliminate low-value features.

\section{METHODS}

Specifically, we conducted a feature ablation study to assess the
informativeness of each feature group and a feature elimination study
to determine the optimal feature sets for classifying Twitter
tweets. We leveraged an existing, annotated Twitter dataset that was
constructed based on a hierarchical model of depression-related
symptoms \cite{Moweryb,moweryd}. The dataset contains 9,473
annotations for 9,300 tweets. Each tweet is annotated as \emph{no
  evidence of depression} (e.g., ``Citizens fear an economic
depression") or \emph{evidence of depression} (e.g., ``depressed over
disappointment").  If a tweet is annotated \emph{evidence of
  depression}, then it is further annotated with one or more
\emph{depressive symptoms}, for example, \emph{depressed mood} (e.g., ``feeling down in the dumps"), \emph{disturbed sleep} (e.g., ``another restless night"), or \emph{fatigue or loss of energy} (e.g., ``the fatigue is unbearable") \cite{Moweryc}. For each class, every annotation (9,473 tweets) is binarized as the positive class e.g., \emph{depressed mood=1} or negative class e.g., \emph{not depressed mood=0}.

\

\subsection{Features}

\begin{figure*}[t!]
\begin{center}
{\includegraphics[width=18.5cm]{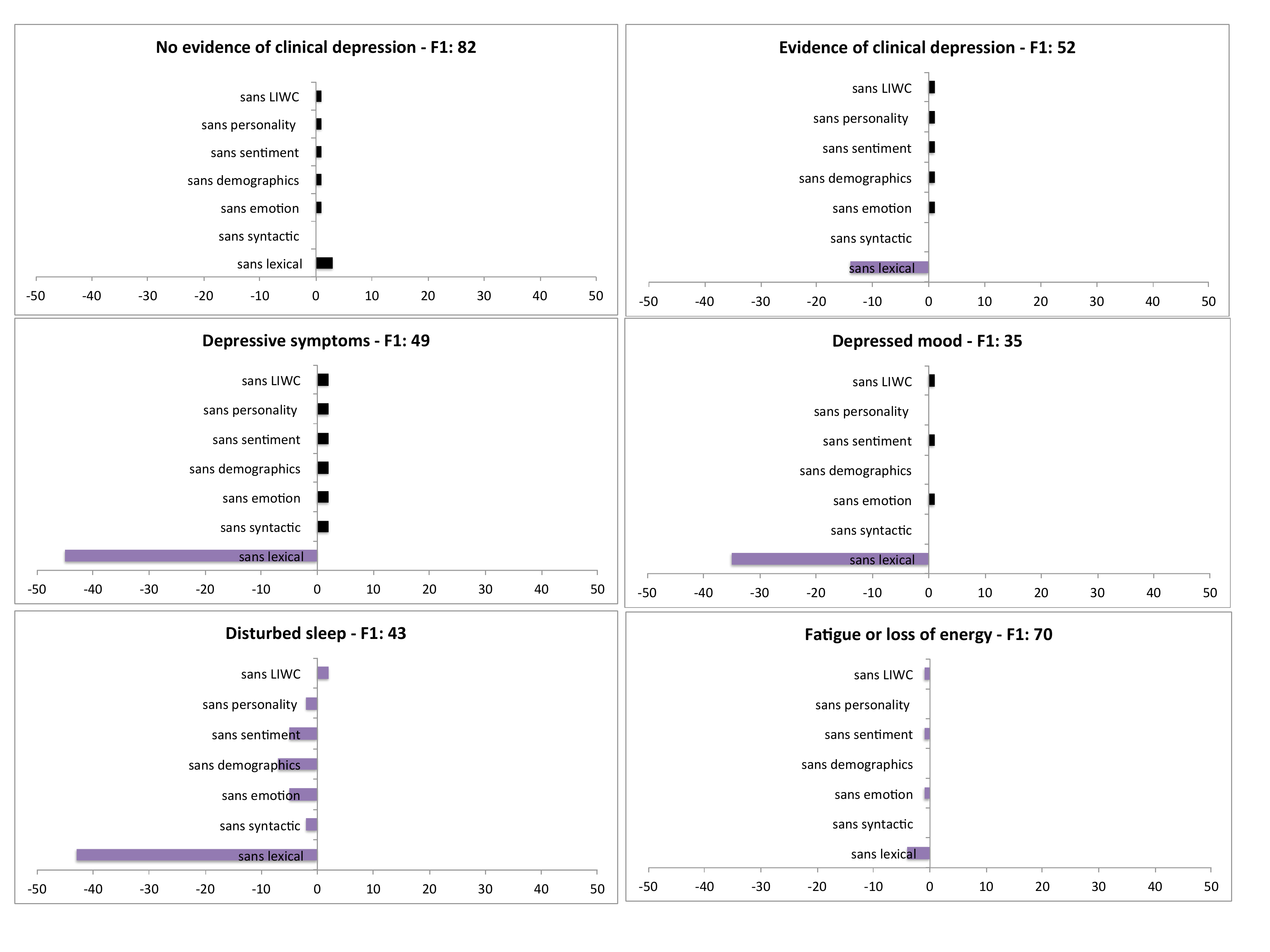}
}
\caption{Feature ablation study: for each class, we plotted the change of average F1-scores from the baseline reported in the titles by ablating each feature set. Black = point gains in F1; Purple = point losses in F1.
}

\label{approx_pr}
\end{center}
\end{figure*}

Furthermore, this dataset was encoded with 7 feature groups with
associated feature values binarized (i.e., present=1 or absent=0) to
represent potentially informative features for classifying
depression-related classes. We describe the feature groups by {\bf type}, subtype, and provide one or more examples of words representing the feature subtype from a tweet:
\\\\

\begin{itemize}
\item {\bf lexical features}, unigrams, e.g., ``depressed'';
\item {\bf syntactic features}, parts of speech, e.g., ``cried'' encoded as V for verb; 
\item {\bf emotion features}, emoticons, e.g., :( encoded as SAD; 
\item {\bf demographic features}, age and gender e.g., ``this semester'' encoded as an
indicator of 19-22 years of age and ``my girlfriend'' encoded as an indicator
of male gender, respectively; 
\item {\bf sentiment features}, polarity and subjectivity terms with strengths, e.g., ``terrible'' encoded as
strongly negative and strongly subjective; 
\item {\bf personality traits}, neuroticism e.g., ``pissed off'' implies neuroticism; \\
\item {\bf LIWC Features}\footnote{Linguistic Inquiry and Word Count \cite{pennebaker2001linguistic}}, indicators of an
individual's thoughts, feelings, personality, and motivations, e.g., 
``feeling'' suggestions perception, feeling, insight, and cognitive
mechanisms experienced by the Twitter user.
\end{itemize}
 A more detailed description of leveraged features and their values, including
LIWC categories, can be found in \cite{Moweryc}.      
  
	Based on our prior initial experiments using these feature groups \cite{Moweryc}, we learned that support vector machines perform with the highest F1-score compared to other supervised approaches. For this study, we aim to build upon this work by conducting two experiments: 1) to assess the contribution of each feature group and 2) to determine the optimal percentile of top ranked features for classifying Twitter tweets in the depression schema hierarchy. \\\\

\subsection{Feature Contribution}
Feature ablation studies are conducted to assess the informativeness of a feature group by quantifying the change in predictive power when comparing the performance of a classifier trained with the all feature groups versus the performance without a particular feature group. We conducted a feature ablation study by holding out (sans) each feature group and training and testing the support vector model using a linear kernel and 5-fold, stratified cross-validation. We report the average F1-score from our baseline approach (all feature groups) and report the point difference (+ or -) in F1-score performance observed by ablating each feature set.

\subsection{Feature Elimination}
Feature elimination strategies are often taken 1) to remove irrelevant or noisy features, 2) to improve classifier performance, and 3) to reduce training and run times. We conducted an experiment to determine whether we could maintain or improve classifier performances by applying the following three-tiered feature elimination approach:
\begin{itemize}
\item \underline{Reduction} We reduced the dataset encoded for each class by eliminating features that occur less than twice in the full dataset.
\item \underline{Selection} We iteratively applied Chi-Square feature selection on the reduced dataset, selecting the top percentile of highest ranked features in increments of 5 percent to train and test the support vector model using a linear kernel and 5-fold, stratified cross-validation.
\item \underline{Rank} We cumulatively plotted the average F1-score performances of each incrementally added percentile of top ranked features. We report the percentile and count of features resulting in the first occurrence of the highest average F1-score for each class.  

\end{itemize}
All experiments were programmed using scikit-learn 0.18\footnote{\url{http://scikit-learn.org/stable/}}.

\section{RESULTS}

From our annotated dataset of Twitter tweets (n=9,300 tweets), we conducted two feature studies to better understand the predictive power of several feature groups for classifying whether or not a tweet contains \emph{no evidence of depression} (n=6,829 tweets) or \emph{evidence of depression} (n=2,644 tweets). If there was evidence of depression, we determined whether the tweet contained one or more \emph{depressive symptoms} (n=1,656 tweets) and further classified the symptom subtype of \emph{depressed mood} (n=1,010 tweets), \emph{disturbed sleep} (n=98 tweets), or \emph{fatigue or loss of energy} (n=427 tweets) using support vector machines.
    From our prior work \cite{Moweryc} and in Figure 1, we report the performance for prediction models built by training a support vector machine using 5-fold, stratified cross-validation with all feature groups as a baseline for each class. We observed high performance for \emph{no evidence of depression} and \emph{fatigue or loss of energy} and moderate performance for all remaining classes.

\subsection{Feature Contribution}

By ablating each feature group from the full dataset, we observed the following count of features - {\bf sans lexical}: 185, {\bf sans syntactic}: 16,935, {\bf sans emotion}: 16,954, {\bf sans demographics}: 16,946, {\bf sans sentiment}: 16,950, {\bf sans personality}: 16,946, and {\bf sans LIWC}: 16,832.
	In Figure 1, compared to the baseline performance, significant drops in F1-scores resulted from {\bf sans lexical} for \emph{depressed mood} (-35 points), \emph{disturbed sleep} (-43 points), and \emph{depressive symptoms} (-45 points). Less extensive drops also occurred for \emph{evidence of depression} (-14 points) and \emph{fatigue or loss of energy} (-3 points). In contrast, a 3 point gain in F1-score was observed for \emph{no evidence of depression}. We also observed notable drops in F1-scores for \emph{disturbed sleep} by ablating {\bf demographics} (-7 points), {\bf emotion} (-5 points), and {\bf sentiment} (-5 points) features. These F1-score drops were accompanied by drops in both recall and precision. We found equal or higher F1-scores by removing non-{\bf lexical} feature groups for \emph{no evidence of depression} (0-1 points), \emph{evidence of depression} (0-1 points), and \emph{depressive symptoms} (2 points).

\subsection{Feature Elimination}
	The initial matrices of almost 17,000 features were reduced by eliminating features that only occurred once in the full dataset, resulting in 5,761 features. We applied Chi-Square feature selection and plotted the top-ranked subset of features for each percentile (at 5 percent intervals cumulatively added) and evaluated their predictive contribution using the support vector machine with linear kernel and stratified, 5-fold cross validation.

In Figure 2, we observed optimal F1-score performance using the following top feature counts: \emph{no evidence of depression}: F1: 87 (15th percentile, 864 features),  \emph{evidence of depression}: F1: 59 (30th percentile, 1,728 features), \emph{depressive symptoms}: F1: 55 (15th percentile, 864 features), \emph{depressed mood}: F1: 39 (55th percentile, 3,168 features), \emph{disturbed sleep}: F1: 46 (10th percentile, 576 features), and \emph{fatigue or loss of energy}: F1: 72 (5th percentile, 288 features) (Figure 1). We note F1-score improvements for \emph{depressed mood} from F1: 13 at the 1st percentile to F1: 33 at the 20th percentile.

\begin{figure*}[t!]
\begin{center}
{\includegraphics[width=18cm]{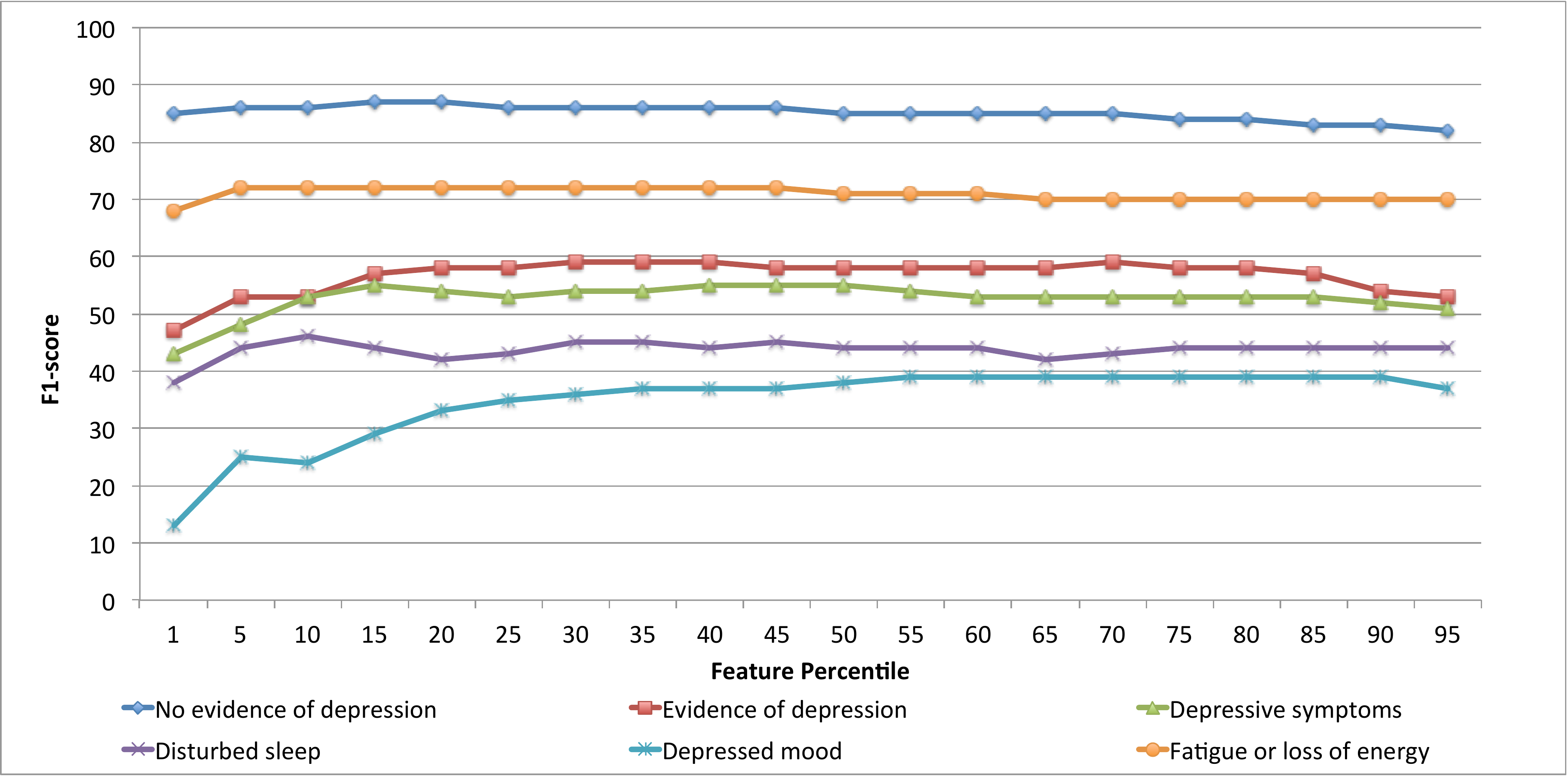}
}

\caption{Feature elimination study: for each class, we plotted the change of average F1-scores for top features of percentiles by adding top-ranked features at 5\% increments to the prediction model. 
}

\label{approx_pr2}
\end{center}
\end{figure*}

\section{Discussion}

We conducted two feature study experiments: 1) a feature ablation study to assess the contribution of feature groups and 2) a feature elimination study to determine the optimal percentile of top ranked features for classifying Twitter tweets in the depression schema hierarchy.

\subsection{Feature Contribution}
Unsurprisingly, {\bf lexical} features (unigrams) were the largest
contributor to feature counts in the dataset. We observed that {\bf
  lexical} features are also critical for identifying \emph{depressive
  symptoms}, specifically for \emph{depressed mood} and for
\emph{disturbed sleep}. For the classes higher in the hierarchy -
\emph{no evidence of depression}, \emph{evidence of depression}, and
\emph{depressive symptoms} - the classifier produced consistent
F1-scores, even slightly above the baseline for \emph{depressive
  symptoms} and minor fluctuations of change in recall and precision
when removing other feature groups suggesting that the contribution of
non-\textbf{lexical} features to classification performance was
limited.  However, notable changes in F1-score were observed for the classes lower in the hierarchy including \emph{disturbed sleep} and \emph{fatigue or loss of energy}. For instance, changes in F1-scores driven by both recall and precision were observed for \emph{disturbed sleep} by ablating {\bf demographics}, {\bf emotion}, and {\bf sentiment} features, suggesting that age or gender (``mid-semester exams have me restless''), polarity and subjective terms (``lack of sleep is killing me''), and  emoticons (``wide awake :('') could be important for both identifying and correctly classifying a subset of these tweets.

\subsection{Feature Elimination}
We observed peak F1-score performances at low percentiles for \emph{fatigue or loss of energy} (5th percentile), \emph{disturbed sleep} (10th percentile) as well as \emph{depressive symptoms} and \emph{no evidence of depression} (both 15th percentile) suggesting fewer features are needed to reach optimal performance. In contrast, peak F1-score performances occurred at moderate percentiles for \emph{evidence of depression} (30th percentile) and \emph{depressed mood} (55th percentile) suggesting that more features are needed to reach optimal performance. However, one notable difference between these two classes is the dramatic F1-score improvements for \emph{depressed mood} i.e., 20 point increase from the 1st percentile to the 20th percentile compared to the more gradual F1-score improvements for \emph{evidence of depression} i.e., 11 point increase from the 1st percentile to the 20th percentile. This finding suggests that for identifying \emph{depressed mood} a variety of features are needed before incremental gains are observed.

\section{Future Work}
Our next step is to address the classification
of rarer depressive symptoms suggestive of major depressive disorder
from our dataset and hierarchy including  \emph{inappropriate guilt},
\emph{difficulty concentrating}, \emph{psychomotor agitation or
  retardation}, \emph{weight loss or gain}, and \emph{anhedonia}
\cite{dsm4,dsm5}. We are developing a population-level
monitoring framework designed to estimate the prevalence of depression (and
depression-related symptoms and psycho-social stressors) over
millions of United States-geocoded tweets. Identifying the most
discriminating feature sets and natural language processing
classifiers for each depression symptom is vital for this goal.

\section{Conclusions}
In summary, we conducted two feature study experiments to assess the contribution of feature groups and to determine the optimal percentile of top ranked features for classifying Twitter tweets in the depression schema hierarchy. From these experiments, we conclude that simple lexical features and reduced feature sets can produce comparable results to the much larger feature dataset.

\section{Acknowledgments}

Research reported in this publication was supported by the National
Library of Medicine of the [United States] National Institutes of
Health under award numbers K99LM011393 and R00LM011393.  This study
was granted an exemption from review by the University of Utah
Institutional Review Board (IRB 00076188). Note that in order to
protect tweeter anonymity, we have not reproduced tweets
verbatim. Example tweets shown were generated by the researchers as
exemplars only.   Finally, we would like to thank the anonymous
reviewers of this paper for their valuable comments.  \\\\\\\\

%
\bibliographystyle{abbrv}
\bibliography{depression_corpus_paper} 

\end{document}